\documentclass[
reprint,
 aps,
 superscriptaddress,
]{revtex4-2}

\usepackage{physics}
\usepackage{dsfont}
\usepackage{color,newfloat}
\usepackage{graphicx}
\usepackage{dcolumn}
\usepackage{bm}
\usepackage{amsmath,amssymb,amsthm}
\usepackage{mathtools}
\usepackage{multirow}
\usepackage[dvipsnames]{xcolor}
\usepackage{hyperref}
\hypersetup{
    colorlinks=true,
    linkcolor=Maroon,
    citecolor=Maroon,
    urlcolor=Maroon
}
\usepackage{siunitx}

\begin{document}


\title{
Deterministic photon source interfaced with a programmable \\silicon-nitride integrated circuit
}

\author{Ying Wang}
\author{Carlos F.D. Faurby}
\affiliation{Center for Hybrid Quantum Networks (Hy-Q), Niels Bohr Institute, University of Copenhagen, Blegdamsvej 17, DK-2100 Copenhagen, Denmark}

\author{Fabian Ruf}
\affiliation{Department of Electrical and Computer Engineering, Aarhus University, 8200 Aarhus N Denmark}

\author{Patrik I. Sund}
\author{Kasper H. Nielsen}
\affiliation{Center for Hybrid Quantum Networks (Hy-Q), Niels Bohr Institute, University of Copenhagen, Blegdamsvej 17, DK-2100 Copenhagen, Denmark}

\author{Nicolas Volet}
\author{Martijn J.R. Heck}
\affiliation{Department of Electrical and Computer Engineering, Aarhus University, 8200 Aarhus N Denmark}

\author{Nikolai Bart}
\author{Andreas D. Wieck}
\author{Arne Ludwig}
\affiliation{Lehrstuhl f{\"u}r Angewandte Festk{\"o}rperphysik, Ruhr-Universit{\"a}t Bochum, Universit{\"a}tsstrasse 150, D-44780 Bochum, Germany}

\author{Leonardo Midolo}

\author{Stefano Paesani}
\email{stefano.paesani@nbi.ku.dk}

\author{Peter Lodahl}
\email{lodahl@nbi.ku.dk}
\affiliation{Center for Hybrid Quantum Networks (Hy-Q), Niels Bohr Institute, University of Copenhagen, Blegdamsvej 17, DK-2100 Copenhagen, Denmark}


\begin{abstract}
We develop a quantum photonic platform that interconnects a high-quality quantum dot single-photon source and a low-loss photonic integrated circuit made in silicon nitride.
The platform is characterized and programmed to demonstrate various multiphoton applications, including bosonic suppression laws and photonic entanglement generation.
The results show a promising technological route forward to scale-up photonic quantum hardware. 
\end{abstract}

\date{\today}

\maketitle



\section{Introduction}

Single photons are a key enabler in emerging quantum technologies including secure quantum communications~\cite{kolodynski2020device, kimble2008quantum} and quantum computing~\cite{Knill2001opticalquantumcomputing, Pan2020bosonsampling, madsen2022quantum}. 
A central challenge in developing scalable photonic quantum hardware is to realize high-quality single-photon sources (SPSs) and interface them with advanced photonic integrated circuits (PICs).
Single-photon sources based on solid-state quantum emitters, such as quantum dots (QDs) or color defect centers, are suitable platforms to deterministically generate near-ideal single photons~\cite{lodahl2022deterministic,aharonovich2016solid}. 
These high-quality photon sources are embedded in III-V semiconductor materials like  gallium arsenide (GaAs) or diamond, and large-scale PICs have not yet been realized on these material platforms since fabrication processes are relatively immature. 
Consequently, a hybrid approach is favorable where the photon source chip is feeding a PIC fabricated on a mature material platform to realize a multi-chip photonic quantum processor~\cite{uppu2021quantum}. 
Ultimately the sources may be heterogeneously integrated on the PIC platform ~\cite{davanco2017heterogeneous,shadmani2022integration,osada2019strongly,wan2020large}.
To build a scalable quantum photonic platform, multiple key requirements need to be simultaneously fulfilled. 
The source must generate highly pure (i.e., emit only one photon at a time) and highly indistinguishable photons as generally required for quantum information processing.
Subsequently, the inevitable loss associated with coupling between the source chip and the processor PIC should be reduced in order to eventually scale up the technology to many photons.
Finally, the PIC chip needs to be low-loss at the operation wavelength of the photon source. 

\begin{figure*}
    \centering
    \includegraphics[width=\textwidth]{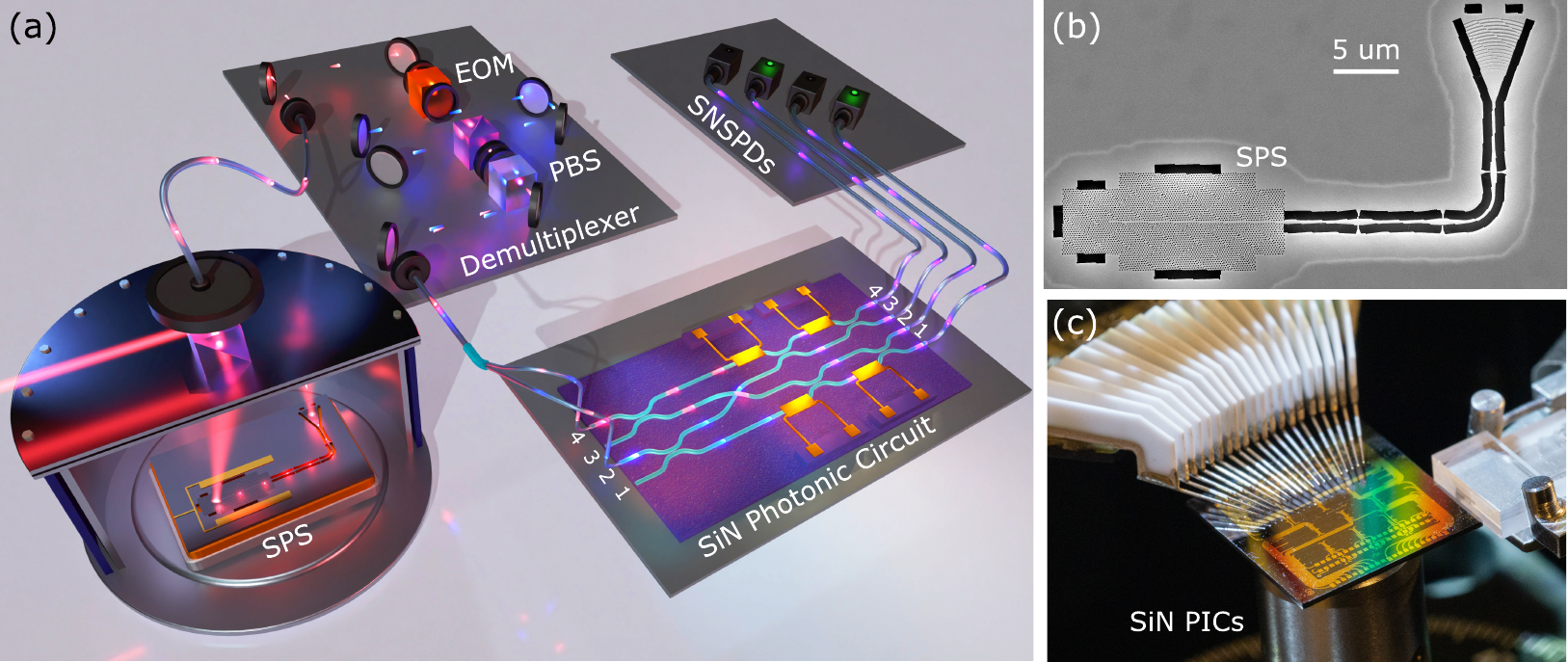}
    \caption{(a) Experimental set-up overview. A QD embedded in a planar nanophotonic chip is excited within a cryostat to emit a stream of single photons (lower left). The stream of single photons is then converted into two streams of simultaneous photons via a free space two-mode demultiplexer (upper left), which relies on a resonant electro-optic modulator (EOM) and a polarizing beams splitter (PBS). The photon streams are coupled ins a programmable SiN PIC operating at room temperature (lower right) and the output is measured via superconducting-nanowire-single-photon-detectors (SNSPDs) (upper right) 
    (b) Scanning electron microscope (SEM) image of the GaAs nanophotonic device.
    (c) Image of the SiN device and set-up.
    }
    \label{fig:1}
\end{figure*}
In this work, we report on the development of a modular  photonic platform with all the key functionalities described above. 
We demonstrate the coupling of single photons from an InGaAs  QD embedded in a planar GaAs photonic-crystal waveguide to a programmable low-loss PIC in SiN. 
Being a mature and CMOS-compatible platform, SiN is highly promising for developing the scalable and low-loss PICs required for photonic quantum technologies, with complex programmable circuits~\cite{taballione2019, arrazola2021}, losses as low as $1$~dB/m~\cite{chanana2022ultra}, and  cryogenic operation~\cite{dong2022} recently demonstrated. 
Importantly, SiN is also highly transparent at the typical emission wavelengths of QDs and enables heterogeneous integration~\cite{davanco2017heterogeneous, wan2020large,chanana2022ultra}, although processing of photons from QD sources has yet to be demonstrated on this platform.
We have realized this and report on high-fidelity multiphoton on-chip operations with a programmable SiN PIC.   
These capabilities are showcased via exemplary photonic quantum information protocols: tests of bosonic suppression laws, i.e., experiments on generalized Hong-Ou-Mandel interference, and the probabilistic generation of  entangled photonic Bell states.

\begin{figure*}
    \centering
    \includegraphics[width=0.9\textwidth]{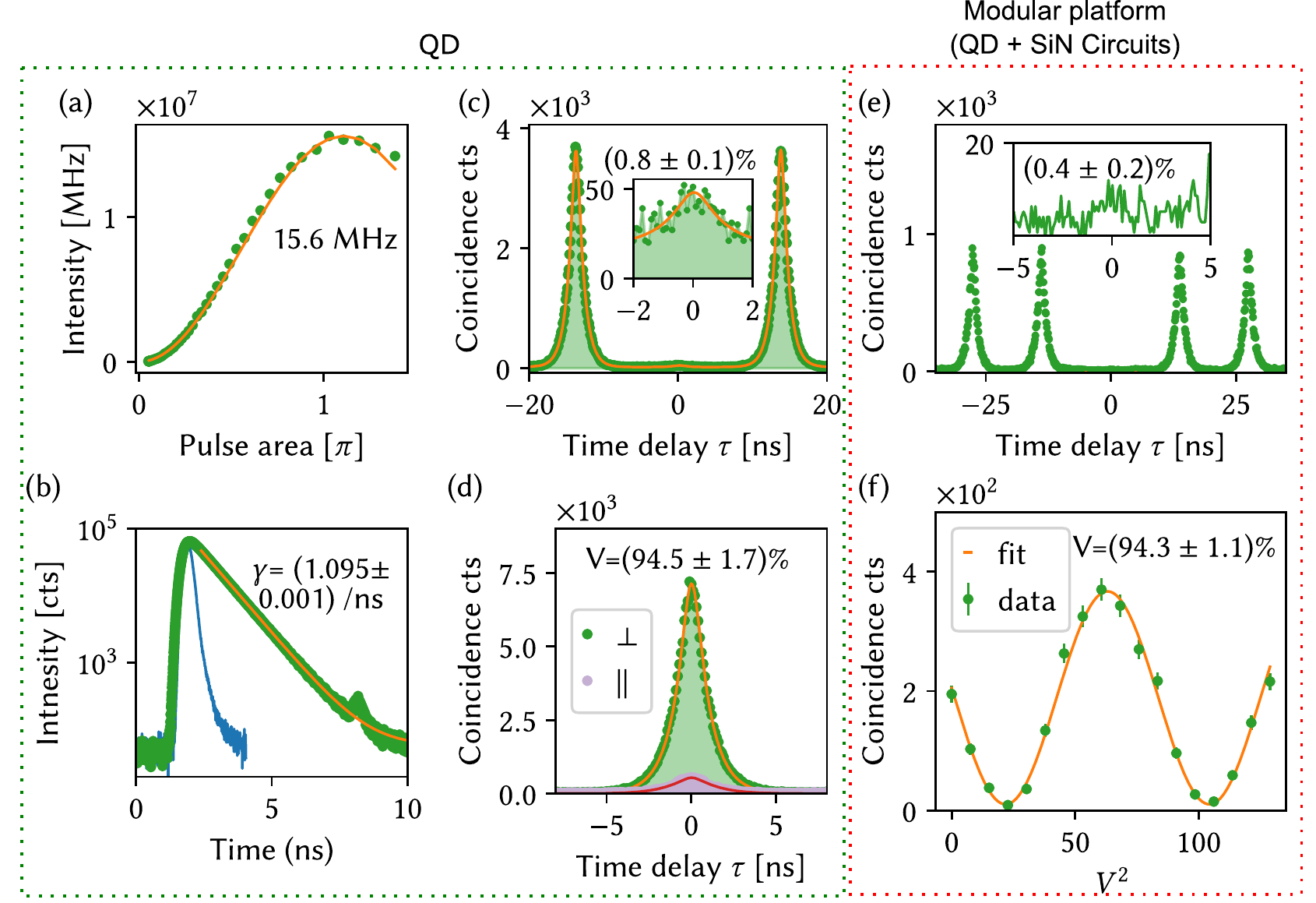}
    \caption{
    Single-photon source characterization performed with free-space optics (left) and on-chip using the SiN PICs (right). 
    (a) Resonance fluorescence data showing the onset of Rabi oscillations when resonantly pumping the QD with a laser of pulse duration 18 ps. 
    The QD is biased at a fixed external voltage of 1.256 {V} at a wavelength of 942\,nm. 
    (b) Time-resolved fluorescence measurement of the spontaneous emission rate of the emitter (green data), modeled with a single exponential decay (orange curve). 
    The blue curve represents the characterized intrinsic  instrument response function from the excitation laser pulse. 
    (c) Single-photon purity $g^{(2)}$(0) characterization performed with a free-space Hanbury Brown and Twiss (HBT) setup, and (e) using the SiN PIC.
    Indistinguishability data obtained (d) in the free-space setup and (f) using an on-chip MZI circuit. In all figures, data are shown as dots and numerical fittings are shown as solid lines (see Supplementary Information for details on the analysis~\cite{SI}).
    }
    \label{fig:2}
\end{figure*}

\section{Platform overview and experimental setup}

The experimental setup for the developed modular platform is shown in Fig.~\ref{fig:1}(a).
We adopt InAs QDs embedded in a p-i-n GaAs diode heterostructure as solid-state photon emitters~\cite{uppu2020scalable}. 
The heterostructure is grown with molecular beam epitaxy, and a distributed Bragg reflector (DBR) is implemented below sacrificial layer ~\cite{SI}. The QDs are flushed enabling to suppress the wetting layer luminescence  for increased emission purity ~\cite{ludwig2017ultra,lobl2019excitons}. Moreover, the QD density is modulated by gradient growth of the  buffer layer thickness~\cite{bart2022wafer}.
High-quality metal gates are realized to quench the charge noise and to tune the emitter emission wavelength via a Stark shift. 
A photonic crystal waveguide (PCW) planar structure, shown in Fig.~\ref{fig:1}(b), is used to interface with the QDs, enabling efficient optical coupling via a combination of radiative leakage suppression and Purcell enhancement featuring broadband operation~\cite{javadi2018numerical}.
A photonic crystal mirror terminates one side of the PCW to enable single-sided emission from the device, see Fig.~\ref{fig:1}(b).
The sample is mounted in a closed-system cryostat at 1.6 K.
A stretched pulsed laser (pulse duration 18 ps, repetition rate of 72 MHz and wavelength of 942\,nm) is used to coherently excite the QD source whereby a neutral exciton in the QD is resonantly excited and subsequently emits triggered single photons. Our implementation closely follows the work of Uppu et al. \cite{uppu2020scalable}, and the relevant differences are discussed in the Supplementary Information \cite{SI}.
The emitted photons are routed to a shallow-etched grating (SEG) for collection.
To achieve high collection efficiency,  a DBR layer is grown below the waveguide to enhance the out-of-plane vertical emission.  
We designed and implemented SEGs that exploit the DBR to boost the reflection of downward emission to significantly enhance the off-plane collection efficiency from the integrated waveguide (see Supplementary Information~\cite{SI} for details on the design and fabrication of the SEGs). 
We measure the efficiency to be 21.5\%, corresponding to a detected 15.6 MHz photon rate (see Fig.~\ref{fig:2}(a)). This efficiency is mainly limited by the not fully optimized coupling from the SEG to the fiber and by the limited efficiency of the etalon used for phononic-sideband filtering, and can readily be further improved. Notably the demonstrated source efficiency is the highest reported to date on planar GaAs nanostructures~\cite{uppu2020scalable}, which enables the integration of the source with the SiN PIC. For a detailed description of the source efficiency, see the Supplementary Information \cite{SI}.

\begin{figure*}
    \centering
    \includegraphics[width=\textwidth]{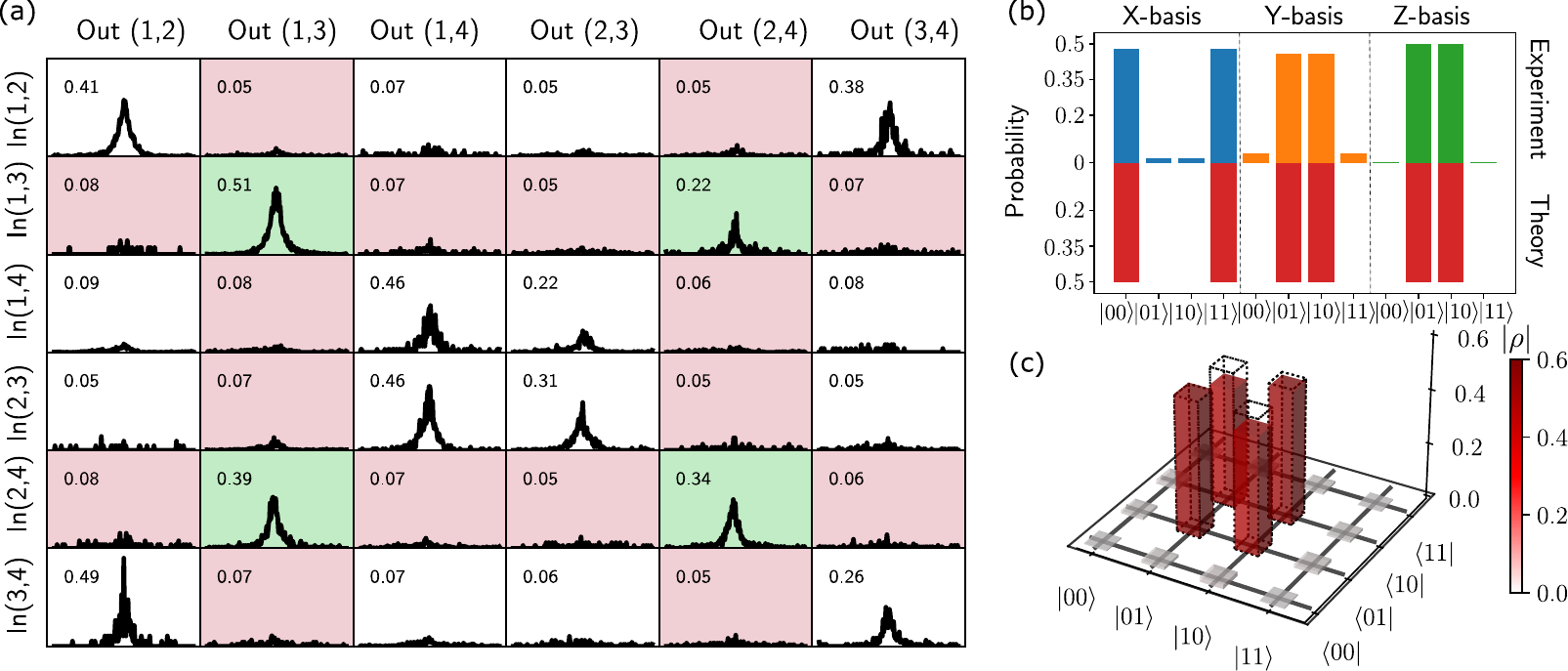}
    \caption{(a) Experiment result of Bosonic suppression laws. Raw detection histograms are reported for all anti-bunched input-output configurations, together with the associated measurement probabilities obtained. The combinations that are expected to be suppressed and enhanced by the suppression law are highlighted in red and green, respectively. (b) Correlation data for Pauli $X_1X_2$, $Y_1Y_2$, and $Z_1Z_2$ qubit measurements on the generated Bell state $\ket{\psi^+}$. Top histograms are measured expectation values, while bottom values are theoretical expectations. (c) Density matrix reconstruction of the post-selected Bell state via state tomography. The red pillars represent the absolute values of the reconstructed density matrix, with transparent dashed pillars representing theoretical expectations. The corresponding measured state fidelity for $\ket{\psi^+}$ is 92\%.}
    \label{fig:3}
\end{figure*}

The photon-emitter platform is optically interfaced to the PIC through optical fibers, as shown in Fig.~\ref{fig:1}(a).
Low-loss photonic circuits are realized on the SiN platform, with Si$_3$N$_4$ waveguides consisting of a rectangular cross-section (100\,nm thick,100\,nm wide) surrounded by a SiO$_2$ cladding. 
Coupling to and from the chip is implemented using a fiber array and on-chip waveguide tapers, as shown in Fig.~\ref{fig:1}(c).
To develop photonic circuitry compatible with the QD source, we designed and optimized components required for programmable photonic quantum circuits tailored to the QD emission wavelength of 940\,nm, and in particular directional couplers for low-loss 50:50 splitters (see Supplementary Information~\cite{SI} for details). 
These SiN components are fabricated in a multi-project wafer run, and compatible with complementary
metal-oxide-semiconductor (CMOS) foundry manufacturing processes, allowing for efficient and scalable fabrication of complex PICs.
In addition to smaller structures used for preliminary characterization of the platform, we implement the reconfigurable four-mode circuit shown in Fig.~\ref{fig:1}(a) which constitutes a tunable Fourier transform interferometer.
This type of circuit finds broad use in photonic quantum information processing~\cite{Tichy2014, Crespi2016, Paesani2021, bacco2021, Marshall2022, bell2022}, and we use it to explore various classes of experiments as reported in the next sections.
The propagation loss in the single-mode SiN waveguides is estimated via the cut-back method~\cite{Peczek2020} to be approximately $0.3$~dB/cm at the QD emission wavelength.
The total coupling losses to the PIC were measured to be between $4.25$ and $7$~dB at the QD emission wavelength (see Supplementary Information ~\cite{SI}). 
This loss value was due to the fabricated devices having an optimal wavelength (with significantly reduced losses of $\sim2$ dB) different from the targeted one, and can be readily improved by adjusting the taper width at the facet. 
To perform multi-photon on-chip experiments, we add a free-space demultiplexer setup, as shown in Fig.~\ref{fig:1}(a), to spatially separate and temporally synchronize consecutive photons emitted by the QD, yielding simultaneous photons pairs at two different input modes of the PIC.

The single-photon source was initially characterized in a free-space setup. The lifetime of the neutral exciton was measured to be 917 {ps} (see Fig.~\ref{fig:2}(b)). A near-unity single-photon purity of 99.2\% at $\pi$ pulse excitation (see Fig.~\ref{fig:2}(c)) was recorded together with an intrinsic Hong-Ou-Mandel (HOM) interference of subsequently emitted photons of $V_\text{HOM} =(94.5 \pm$ 1.7)\% (see Fig.~\ref{fig:2}(d)) after sideband filtering with an etalon. This corresponds to a pure dephasing rate of $0.03 \text{ns}^{-1}$~\cite{tighineanu2018phonon}.
To characterize the quality of the PIC, we test how these near-ideal properties are affected by the circuit.   
To do so, we characterize the on-chip photon purity by implementing an integrated Hanbury-Brown and Twiss 
experiment where photons from the QD are directly injected into the SiN PIC and routed to a Mach-Zehnder interferometer (MZI) with the internal phase difference set to $\pi/2$ to implement a 50:50 beam-splitter. 
$g^{(2)}(0)$ is measured by recording coincidences between the two outputs normalized by coincidences between different pulses, as shown in Fig. \ref{fig:2}(e), obtaining a purity of 99.6\%, 
which is compatible with the free-space value.
To perform a HOM experiment of the on-chip photon indistinguishability, we use the demultiplexer to send single photons into each of the two input ports of the MZI structure,
while scanning the internal phase shifter and coincidences between the two output detectors were recorded. 
By scanning the internal phase shifter of the MZI, we obtain the HOM quantum interference fringe shown in Fig.~\ref{fig:2}(f), from which the corresponding HOM visibility is extracted~\cite{adcock2019programmable}.
We find $V_\text{HOM}=(94.3\pm 1.1)\%$ in full agreement with the free-space value, indicating that no degradation of the photon properties was observed due to the PIC platform.     

\section{Bosonic suppression laws}

To investigate the performance of our platform for practical photonic quantum information processing applications, we use it to test generalized HOM quantum interference in the form of bosonic suppression laws in a discrete Fourier transform (DFT) interferometer. 
Introduced by Tichy et al.~\cite{Tichy2010, Tichy2014, Crespi2016}, they
are a generalization of the HOM quantum interference, which is a suppression of coincidences when photons are interfered on a balanced beam
splitter (two-mode DFT), to higher number of modes and photons.
Applications of bosonic suppression laws in photonic quantum technologies include the verification of quantum advantage experiments via boson sampling~\cite{Tichy2014, Crespi2016} and schemes for high-dimensional photonic quantum computing~\cite{Paesani2021}.
By programming the phase shifters in the SiN PIC shown in Fig.~\ref{fig:1}, we implement a four-mode DFT described by the unitary transformation matrix $U_{j,k} = \exp[(j-1)(k-1) 2\pi \text{i}/4]/2$, where $j,k \in \{1, 2, 3, 4\}$ label the circuit spatial modes (see Supplementary Information for circuit details~\cite{SI}).
The suppression law states that whenever a cyclic input configuration of $n$ photons is injected to the DFT all output configurations where the value of $\sum_{i=1}^n (c_i \mod{n})$ is different than zero are suppressed (i.e., have zero amplitude), where $c_i$ represents the output mode number of the $i$-th photon~\cite{Tichy2010}.
For example, considering $n=2$ photons in a four-mode DFT, a cyclic input configuration could be sending two photons in the input modes $(1,3)$, for which suppression is expected at the output mode configurations $(c_1=1, c_2=2)$, $(c_1=1, c_2=4)$, $(c_1=2, c_2=3)$ and $(c_1=3, c_2=4)$ such that $c_1 + c_2$ is an odd number.
We experimentally test the suppression law by demultiplexing photon pairs from the quantum emitter and injecting them in all six possible pairs of different input modes of the integrated DFT.
The output statistics for all antibunched two-photon output configurations are measured via four superconducting nanowire single-photon detectors (SNSPDs) connected to the output modes (see Fig.~\ref{fig:1}) and analyzed with a time-tagger.
The results are shown in Fig.~\ref{fig:3}(a), where the expected suppressed and enhanced configurations are highlighted in red and green, respectively. 
The experimental data show good agreement with the theoretical expectations, i.e. showing clear suppression of the configurations which sum to an odd number for the cyclic inputs. 
The remaining contributions to the amplitudes of suppressed configurations can be well explained considering the finite degree of indistinguishability of the photons generated by the quantum emitter, as analyzed in the Supplementary Information~\cite{SI}. 

\section{Photonic entanglement generation and characterization}
We now proceed to investigate the generation of photon--photon entanglement, a key building block in photonic quantum information processing. 
A pair of path-encoded qubits can be encoded in two photons propagating in the four-mode SiN PIC associating the occupation of each mode to the computational qubit states as $1 \mapsto \ket{0}_1$, $2 \mapsto \ket{1}_1$, $3 \mapsto \ket{0}_2$, and $4 \mapsto \ket{1}_2$.
When we inject a photon in each of the input modes 1 and 3, the SiN circuit transforms the input state into a superposition, which leads to the entangled Bell state $\ket{\psi^+}=(\ket{0}_1\ket{1}_2+\ket{1}_1\ket{0}_2)/\sqrt{2}$. This state is obtained after post-selecting on the two-photon coincidence events where each photon ends up in separate qubit modes ~\cite{Ralph2002, Pont2022fourphotonGHZ, fyrillas2023certified}, which happens with a success probability of $50\%$.
The MZIs and phase-shifters at the output of the circuit (see Fig.~\ref{fig:1}) can be used to measure the two qubits in arbitrary qubit bases, which we use to perform a tomographic reconstruction and analysis of the generated entangled state.
More details on the functioning of the circuit can be found in the Supplementary Information~\cite{SI}.
The experimental results are reported in Fig.~\ref{fig:3}(b)-(c).
The two input photons are again obtained by demultiplexing the stream of photons from the QD.
Correlation data in the Pauli matrices bases $X_1X_2$, $Y_1Y_2$, and $Z_1Z_2$ are shown in Fig.~\ref{fig:3}(b), showing high correlations in the first case and anticorrelations for the latter two, as expected for the $\ket{\psi^+}$ state.
Measuring the qubits in other additional Pauli bases allows us to reconstruct the density matrix of the output state via quantum state tomography~\cite{James2001}.
The obtained density matrix is reported in Fig.~\ref{fig:3}(c), which presents a fidelity of 92\% with the ideal $\ket{\psi^+}$ entangled state, providing another confirmation of the good performance of the PIC. 
The remaining infidelity can again be attributed to partially distinguishable photons, as analyzed in the Supplementary Material~\cite{SI}.

\section{Conclusion}

We have demonstrated a modular quantum photonic platform based on a high-quality and efficient QD single-photon source interfaced with a low-loss and scalable SiN PIC. 
This allowed  us to demonstrate  various on-chip multiphoton quantum experiments relevant to emerging photonic quantum technologies.
The Bosonic suppression laws were realized for the first time for a solid-state photon source, which was enabled by the controllable integrated circuits. 
Several improvements of the experiments are readily possible by further optimization of the components design and improved packaging of the chips in order to reduce coupling losses. 
This will enable experiments  with an increasing number of photons and ultimately large-scale photonic quantum technologies. 
The direct heterogeneous integration of the solid-state photon sources with SiN PICs represents a potentially important route to further reduce losses.
For example, recent advances have shown the possibility of large-scale integration of quantum emitters and waveguides with pick-and-place techniques in similar platforms~\cite{wan2020large, chanana2022ultra}.
Such techniques can be readily adapted also to our high-quality photon sources as they are embedded in photonic nanostructures suitable for transfer-printing onto SiN~\cite{roelkens2022micro}. 
Achieving a platform that explores both state-of-the-art QD photon sources with the scalability of CMOS-compatible PICs in SiN  represents a promising avenue for developing photonic quantum technologies.
In this work, we have taken the first steps to explore such opportunities.

\noindent\textbf{Acknowledgments}
We thank Freja T. Østfeldt, Cecile T. Olesen, Camille Papon, Søren Preisler, and Mikkel T. Mikkelsen for experimental assistance. We acknowledge LioniX International B.V. for the SiN device fabrication.
We acknowledge Larissa Vertchenko for assistance with producing Fig.~\ref{fig:1} (a) and Julian Curry Robinson-Tait for taking the photograph shown in Fig.~\ref{fig:1} (c). We acknowledge funding from the Danish National Research Foundation (Center of Excellence “Hy-Q,” grant number DNRF139), the Novo Nordisk Foundation (Challenge project "Solid-Q"), and Innovationsfonden (grant No. 9090-00031B, FIRE-Q). 
S.P. acknowledges funding from the Cisco University Research Program Fund (nr. 2021-234494), from the Marie Skłodowska-Curie Fellowship project QSun (nr. 101063763), and from the VILLUM FONDEN research grant VIL50326.
L. M. acknowledges funding from the European Research Council (ERC) under the European Union's Horizon 2020 research and innovation program (No. 949043, NANOMEQ).
N.B. A.D.W. and A.L. acknowledge funding from the BMBF contract (No. 16KISQ009).
%


\bibliography{bib.bib}

\end{document}